\documentclass[aps,prb,a4paper,preprint,showkeys,showpacs,floatfix]{revtex4}

\usepackage{graphicx}
\usepackage{dcolumn}
\usepackage{bm}
\usepackage{amsmath}
\usepackage{amssymb}

\begin{document}


\title{\emph{Ab initio} investigation of intermolecular
interactions in solid benzene}

\author{O. Bludsk\'{y}}
\email{ota.bludsky@uochb.cas.cz}
\author{M. Rube\v{s}}
\affiliation{
Institute of Organic Chemistry and Biochemistry, Academy of Sciences of the
Czech Republic  \\ and Center for Biomolecules and Complex Molecular Systems,
Flemingovo n\'{a}m. 2, 166 10 Prague 6, Czech Republic
}

\author{P. Sold\'{a}n\footnote{the corresponding author}}
\email{pavel.soldan@mff.cuni.cz}
\affiliation{
Charles University in Prague, Faculty of Mathematics and Physics, Department of
Chemical Physics and Optics, Ke Karlovu 3, 121 16 Prague 2, Czech Republic
}

\date{\today}

\begin{abstract}
A computational strategy for the evaluation of the crystal lattice constants and cohesive energy of the weakly bound molecular solids is proposed.
The strategy is based on the high level \textit{ab initio} coupled-cluster determination of the pairwise additive contribution to the interaction energy. The zero-point-energy correction and non-additive contributions to the interaction energy are treated using density functional methods. The experimental crystal lattice constants of the solid benzene are reproduced, and the value of 480 meV/molecule is calculated for its cohesive energy.
\end{abstract}

\maketitle

\section{\label{sec:introduction}INTRODUCTION}

In the last 50 years, weakly bound
solids, including rare-gas solids and organic molecular solids, have been the subject of considerable research interest.\cite{Pollack:1964,Apkarian:1999,Schwoerer:2006}
Many experimental measurements and theoretical predictions of the crystal structures, lattice constants, cohesive energies,
lattice dynamics and phase behavior of weakly bound solids have been published.
While experiments can be performed directly on the weakly bound solids, theoretical predictions are
often derived from models that build upon pairwise additive interaction
energies and include the most important refinements such as non-additive contributions and zero-point energy.
Such information is obtained from calculations on the corresponding weakly bound dimers, trimers or tetramers.
Therefore, a good knowledge of the intermolecular interactions is essential for the reliable theoretical determination
of the properties of the weakly bound solids.

Recently, a series of the state-of-the-art theoretical studies on rare-gas solids has been performed
using the high-level \emph{ab initio} coupled-cluster method.\cite{Stoll:1999,Stoll:2000,Paulus:2002,Schwerdt:2006}
In these studies, the pairwise additive contributions to the interaction energy were represented by empirical potentials, the zero-point-energy (ZPE) corrections were calculated within the harmonic approximation from the pairwise additive contributions, and non-additive contributions
were determined using coupled-cluster calculations with single and double excitations and a perturbative
treatment of the triple excitations (CCSD(T)).\cite{Cizek:ACP1969} It was revealed that theoretical predictions of the structural parameters
at this level of theory agree with experiment, well within the experimental error
bars.

It would be ideal if it were possible to apply the same treatment to weakly bound molecular solids.
However, empirical potentials are not always available. Moreover, when the corresponding molecule is relatively large,
the high level \emph{ab initio} methods, such as CCSD(T), are not always affordable.
That is why density-functional-theory (DFT) calculations are often performed to determine lattice constants and cohesive energies.
On the other hand, predicting the structural parameters of weakly bound molecular crystals using merely the DFT
methodology is not straightforward due to the inability of the DFT to account for the dispersion interactions.

In this paper, we present a combined approach to the weakly bound molecular crystals in which the pairwise additive contribution to the interaction
energy is calculated at the CCSD(T) level and the ZPE
correction and non-additive contributions are treated purely at the DFT level.
We illustrate our approach by calculating the cohesive energy and crystal lattice parameters of solid benzene, the simplest real crystalline
system, in which interactions between aromatic molecules can be studied. In a very recent study on the solid-benzene vibrational dynamics,\cite{Kearley:2006} the Perdew-Burke-Ernzerhof (PBE) exchange-correlation functional\cite{Perdew:PRL1996} performed extremely well,
and its results were in an excellent agreement with the inelastic neutron-scattering spectroscopy data.
For precisely this reason, we have selected the PBE functional as a basic DFT tool for our study. The methodology of our approach is described in Section II. In Section III, the results obtained for the solid benzene are discussed, followed by  conclusions in Section IV.

\section{METHODOLOGY}

In general, the interaction energy of a molecular crystal can be expressed as
\begin{equation}
\label{cohesionexact}
E(\mathbf{a})= E^{(2)}(\mathbf{a}) + \sum_{n\ge 3} E^{(n)}(\mathbf{a}) +  \Delta E_{\rm ZPE}(\mathbf{a}),
\end{equation}
where the 3x3 matrix $\mathbf{a}$ represents lattice parameters, $E^{(2)}$ is a pairwise additive (two-body) contribution,  $\sum_{n\ge3} E^{(n)}$ represents the sum of many-body non-additive terms, and $\Delta E_{\rm ZPE}$  is a zero-point-energy correction.
It has been shown that the periodic DFT methods can accurately describe the crystal structure (mutual orientation of
molecules) and spectra of molecular solids. On the other hand, the DFT with local density functionals badly fails for binding, and
therefore a higher level of theory has to be used for the calculation of cohesive energies and equilibrium lattice constants.
Our computational methodology consists of the following steps: (i) a fixed-volume
plane-wave DFT geometry optimization for the given lattice parameters $\mathbf{a}$, (ii) a calculation of the two-body term using a coupled-cluster method, and (iii) an estimate of the many-body contributions and  a calculation of the ZPE correction at the periodic DFT level.
The pairwise additive contribution $E^{(2)}$ represents by far the largest
contribution to the total interaction energy, which is the reason why this term was evaluated at the highest level of theory feasible for solid benzene.

\subsection{Plane-wave DFT calculation}

The fixed-volume structural optimizations for the solid benzene were performed
with the lattice constants obtained from the one-parameter scaling of the experimentally determined unit cell (\emph{Pbca} orthorhombic cell with $a=7.355$\,\AA, $b=9.371$\,\AA, and $c=6.699$\,\AA).\cite{David:PhysB1992}
The calculations were carried out using the periodic plane-wave DFT with the Perdew-Burke-Ernzerhof (PBE) exchange-correlation functional.\cite{Perdew:PRL1996}
The cut-off energy of 800 eV and hard potentials for carbon and hydrogen atoms (ENMAX=700 eV) were employed.  The Kohn-Sham equations were solved using a plane-wave basis set by the projector-augmented wave (PAW) method of Bl\"{o}chl \cite{Bloechl:PRB1994} as adopted by Kresse and
Joubert.\cite{Kresse:PRB1999} The Brillouin-zone sampling was carefully checked for convergence
with the number of $\mathbf{k}$ points.  The Vienna \textit{Ab initio} Simulation Package (VASP) \cite{Kresse:PRB1993} was used for all
the plane-wave PBE calculations.

\subsection{Calculations of the $E^{(2)}$ term}

The CCSD(T) calculations of the interacting benzene pairs in a crystal
(for intermolecular distances of less than 10\,\AA) were carried out with the augmented
correlation-consistent valence-double-$\zeta$ basis set with polarization functions\cite{Dunning:JCP1989}
(AVDZ) using the MOLPRO 2002.6 program suite.\cite{Molpro}
The complete basis set (CBS) extrapolation was performed using a simple correlation-energy dependence
on the basis-set cardinal number $X$ ($E_{X}=E_{\rm CBS}+AX^{-3}$).\cite{Halkier:CPL1998} The density-fitting
spin-component-scaled MP2 (SCS-MP2) method\cite{Grimme:JCP2003} was employed for the correlation
energy extrapolation (using the AVTZ and AVQZ basis sets). The Hartree-Fock energies were
calculated using the AV5Z basis on the carbon atoms and the V5Z basis on the hydrogen atoms
(the AV5Z$^{*}$ basis set). The CCSD(T)/CBS estimate $E^{\rm CC}_{\rm CBS}$ for the benzene dimer was evaluated
according to the formula
\begin{equation}
\label{CCSDCBS}
E^{\rm CC}_{\rm CBS}= E^{\rm CC}_{\rm AVDZ} - E^{\rm MP2}_{\rm AVDZ} + E^{\rm MP2-HF}_{\rm CBS} + E^{\rm HF}_{\rm AV5Z^{*}}.
\end{equation}
All the calculated interaction energies were corrected for the basis set superposition
error (BSSE) using the counterpoise correction method of Boys and Bernardi.\cite{Boys:MolPhys1970}
The $E^{(2)}$ calculations were performed using the benzene geometry of Gauss and Stanton \cite{Gauss:JPCA2000}
($r_{\rm CC}=1.3915$\,\AA, $r_{\rm CH}=1.0800$\,\AA).

\subsection{Asymptotic quadrupole-quadrupole and dispersion corrections
to $E^{(2)}$}

\begin{figure}
\includegraphics{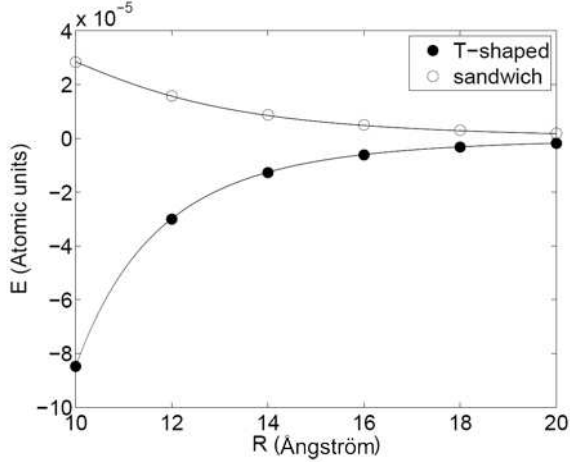}
\caption{Asymptotic part of the intermolecular benzene-benzene potential
calculated for the sandwich and T-shaped structures.\label{fig:asymptfig}}
\end{figure}

An asymptotic intermolecular benzene-benzene potential was derived from the
coupled-cluster data for the sandwich ($D_{6h}$) and T-shaped ($C_{2v}$) structures of the benzene dimer.
The following functional form of the asymptotic behavior was assumed
\begin{equation}
\label{asympt}
E^{\rm asympt} = \frac{\omega\,Q_{zz}^{2}}{R^{5}} + \frac{C_{6}}{R^{6}} + \frac{C_{8}}{R^{8}},
\end{equation}
where $R$ is the distance between the monomer centers of mass, $Q_{zz}$ is an effective (as it includes
higher-order electrostatic interactions) quadrupole moment of the benzene molecule, and $C_{6}$
and $C_{8}$ are isotropic parameters of the dispersion interaction. The
angular factor $\omega$ describes an  angular dependence of the quadrupole-quadrupole
interaction for symmetric top molecules\cite{Stone:1996} ($\omega=-3$ for the T-shaped structure and $\omega=6$ for the
sandwich structure of the benzene dimer). One-dimensional CCSD(T)/AVDZ scans through the benzene-dimer potential energy surface were calculated for six points between 10\,\AA\ and 20\,\AA\ with a step of 2\,\AA\ for both structures (see Fig \ref{fig:asymptfig}). Then the parameters $Q_{zz}$, $C_{6}$, and $C_{8}$ were fitted to the BSSE-corrected CCSD(T)/AVDZ interaction energies of the benzene dimer with the following results: $Q_{zz}=-5.509(5)$ a.u., $C_{6}=-1.66(4)\,10^{3}$ a.u., and $C_{8}=-1.7(2)\,10^{5}$ a.u.  Eq.(\ref{asympt}) was used to calculate
the asymptotic correction to the solid-benzene pairwise additive energy for intermolecular distances larger than 10\,\AA.

\subsection{Plane-wave DFT calculation of many-body and ZPE corrections}

The sum of many-body non-additive terms $\sum_{n\ge3} E^{(n)}$ was approximated as the difference
between the total crystal energy (without ZPE) and the pairwise additive contribution, both calculated at the the PBE/PW-800eV level.
A simple cubic cell with a lattice parameter of 20 \AA\ was employed for the benzene-pair plane-wave calculations. The calculated benzene-pair interaction energies were in excellent agreement with the counterpoise corrected benzene-dimer interaction energies obtained from the PBE/AVQZ calculations.
The pairwise additive contributions for the intermolecular distances larger than 10\,\AA\ were calculated using Eq.(\ref{asympt}) with  $Q_{zz}=-4.85(3)$ a.u. and $C_{6}=-3.87(3)\,10^{2}$ a.u. (with $C_{8}$ set to zero) obtained by fitting the corresponding PBE/AVQZ data.

The ZPE correction $\Delta E_{\rm ZPE}$ was evaluated using
zone-centered ($\Gamma$-point) harmonic frequencies calculated at the PBE/PW-800eV level.

\section{RESULTS AND DISCUSSION}

In Table \ref{tab:cohesive}, the individual contributions to the cohesive energy of solid benzene are presented, including
four approximation levels of the pairwise additive contribution.
By comparing the first two rows, it is evident that the basis-set dependence
has to be considered very carefully. Our CCSD(T)/CBS estimates
can be directly compared with the latest theoretical benchmarks for the benzene dimer.
For example, our binding energy for the T-shaped ($C_{2v}$) benzene-dimer structure overestimates the best theoretical
value, calculated at the QCISD(T)/CBS level,\cite{Janowski:CPL2007} by 0.6\%.
The same error has to be expected for the two-body contribution in Table I, and our value is thus likely to be overestimated
by 4 meV/molecule.
It arises from Table I that when taking into account only the nearest neighbors,
the two-body contribution is underestimated by 12\% even at the CBS limit.
The inclusion of the pairs up to 10 \AA\ explicitly (Eq. \ref{CCSDCBS})) and the long-range pairs asymptotically (Eq. \ref{asympt})) was found to be necessary.

The pairwise additive contribution to the cohesive energy is lowered by the many-body and ZPE-correction contributions thus
leading to our best estimates of
480 meV/molecule (C$_6$H$_6$) and 486 meV/molecule (C$_6$D$_6$) of the solid-benzene cohesive energy. The calculated cohesive energy
compares well with the experimental values of the heat of sublimation of 460-560 meV/molecule.\cite{NIST} Note that by definition the
enthalpy of sublimation is equal to the cohesive energy only at the temperature of absolute zero.

The sum of the many-body and ZPE-correction terms forms 17\% of  the pairwise additive contribution. This is only slightly higher than in the case of the heavier rare gas solids (Ar, Kr, Xe), where the pairwise additive contributions are lowered by many-body and ZPE contributions as well, and the sums of the many-body and ZPE terms are in the range of 11-15\% of the pairwise additive contributions.\cite{Stoll:2000}

\begin{table}
\caption{\label{tab:cohesive} Two-body, many-body, and ZPE contributions to the
total cohesive energy of solid benzene (in meV/molecule).}
\begin{ruledtabular}
\begin{tabular}{lcr}
model & level & $E_{\rm coh}$ \\
\hline
two-body nearest neighbor & CCSD(T)/AVDZ  & 436  \\
& CCSD(T)/CBS                    & 507  \\
two-body $R< 10$ \AA  & CCSD(T)/CBS        & 571  \\
two-body incl. asympt. corr. & CCSD(T)& 592  \\
many-body terms  & PBE/800\,eV & -68  \\
ZPE corr. term\footnote{the value for C$_6$D$_6$ is given in parentheses} & PBE/800\,eV & -44 (-38) \\
best CCSD(T)+PBE estimate & & 480 (486)  \\
\end{tabular}
\end{ruledtabular}
\end{table}

This similarity between solid benzene and heavier rare-gas solids is also apparent
when the role of the individual contributions to the crystal lattice constants is examined. Table
\ref{tab:lattice} contains the ratios of the theoretical lattice constants calculated at various approximation levels
to the experimental ones for rare-gas solids and deuterated solid benzene.
It is obvious that the role of the individual contributions is rather similar, particularly in the case of the heaviest rare-gas solid,
solid xenon.

\begin{table}
\caption{\label{tab:lattice} Two-body, many-body, and ZPE contributions to the
theory/experiment ratio for the fcc crystal lattice constant of rare-gas
solids (derived from Ref.[\onlinecite{Stoll:2000}]) and deuterated solid benzene.}
\begin{ruledtabular}
\begin{tabular}{lccccc}
& Ne & Ar & Kr & Xe & C$_6$D$_6$ \\
\hline
two-body term  & 0.958 & 0.982 & 0.980 & 0.987 & 0.987 \\
incl. many-body terms  & 0.963 & 0.989 & 0.987 & 0.993 & 0.993 \\
incl. ZPE corr. term & 1.001 & 1.000 & 0.993 & 0.997 & 1.000 \\
\end{tabular}
\end{ruledtabular}
\end{table}

The asymptotic many-body contribution to the interaction energy has been approximated only at the DFT level (because it was beyond our computational possibilities to perform a CCSD(T) calculation on the benzene trimer
and tetramer).
From the examination of the results on the deuterated solid benzene (Table
\ref{tab:lattice}), it seems that this approximation was sufficient for the accurate  determination of the crystal lattice constants. However, this might also be caused by a fortuitous cancellation of the terms with different signs, because the asymptotic many-body long-range terms
are strongly geometry-dependent, and so in other weakly bound molecular solids such an approximation may prove to be insufficient.

\section{CONCLUSION}

We have presented a computational strategy for the theoretical determination of the cohesive energy and crystal lattice constants of weakly bound molecular solids.
This strategy is based on the high level \textit{ab initio} (CCSD(T)) calculation of the pairwise additive contribution to the interaction energy, whereas the zero-point-energy correction and non-additive contributions are treated using the DFT (PBE). Having applied this strategy to solid benzene, we have been able to reproduce its experimental lattice parameters as well as predict its cohesive energy (480 meV/molecule for C$_{6}$H$_{6}$ and 486 meV/molecule for C$_{6}$D$_{6}$). Further tests and possible improvements of the proposed computational scheme are desirable, particularly the role of the asymptotic many-body terms in weakly bound molecular solids needs closer attention. More studies on the weakly bound molecular crystals are currently being prepared.

\begin{acknowledgments}
This research was supported by the Grant Agency of the Academy of Sciences of the Czech Republic
(grant No. IAA400550613). OB and MR also acknowledge grant No. LC512 and research project No. Z4 055 0506 of the
Ministry of Education, Youth and Sports of the Czech Republic.
\end{acknowledgments}


\end{document}